\documentclass[12pt]{article}
\usepackage{graphicx}
\newcommand{\noi}{\noindent}
\newcommand{\beq}{\begin{equation}}
\newcommand{\eeq}{\end{equation}}

\title{Dynamical Riemannian Geometry and Plant Growth}
\author{J. Pulwicki\footnote{email:jpulwick@ucalgary.ca}~ and D. Hobill\footnote{email:hobill@phas.ucalgary.ca}
\\ Dept. of Physics and Astronomy, University of Calgary \\
Calgary, Alberta, T2N 1N4, Canada}
\date{\empty}
\begin{document}
\maketitle

\section{Introduction}

Mathematical modeling of biological systems has, in recent years, caught the attention of an eclectic mix of
biologists, mathematicians, computer scientists and physicists. The variety and complexity of problems encountered 
so readily in Nature makes biological models an excellent test-bed for nonlinear dynamics, statistics and a number 
of abstract computational theories. In plant leaves, features such as ruffles, symmetric vein patterns and buckling 
have not been explained at the cellular level, and despite displaying geometric elegance, have only been examined 
mathematically in a handful of descriptive, rather than predictive ways. It is our aim to fully describe the 
structure of plant leaves 
using the techniques of differential geometry, fluid flow mechanics and numerical simulations. 

There are two main assumptions that provide the basis for such an analysis. The first is that the cells that make up a  
plant leaf are vanishingly small compared to the size of the leaf, so that a leaf can be 
described as a 2-dimensional differentiable manifold embedded in a flat, 3-dimensional space. 
The second assumption is that nutrients and hormones are distributed continuously throughout the leaf. Indeed, biological
data supports this assumption, and it has been a key component of previous work done in the study of plant growth \cite{SE}. 
Interpreted mathematically, such an assumption permits the use of flow equations to describe the dynamics of
nutrients, hormones and growth in the leaf.
Combining the two assumptions, the problem of leaf growth can be explored as transport of material
on a curved background space.

Though it is relatively straightforward to choose physical assumptions to work with, we cannot forget that the goal is
to model an intricate living system.  The fact that leaves
are often symmetric in vein patterns and shape is particularly striking, since in physics a symmetry indicates
a conserved quantity. Does symmetry in growth allow for an efficient plant? Furthermore, how does the geometry of 
a leaf affect the flow of nutrients and hormones in a plant, and hence its growth? Indeed, these types of 
questions require the factual knowledge of biology as well as insights into the geometric structure of plants.

Even from these simple considerations, we already see the purpose of this work is two-fold. On the one hand, 
we have started with a simple, unanswered
question in biology that might well be asked by a child: why are leaves shaped the way they are? Our approach, 
however, incorporates mathematics that is common to many physical systems but in itself is only partially understood. 
By having a physical system to emulate using our equations, we hope to not only discover how geometry presents
itself in Nature, but also gain a better understanding of the equations that scientists in a variety of disciplines
use today.

\section{The Geometry and Growth of Plants}

Interest in plant geometry and growth is by no means a new field of study. Scientists have tried to quantify both
how plants grow and what is is that might determine their geometry 
for the past century, with efforts coming from three main directions: computer science, quantitative biology and
structural mechanics.  The approaches to understanding plant geometry range
from static models involving single rule based models to analogues with
mechanical systems.  Similarly models of growth can focus on the cellular
scale to the organism over all where continuum systems are applied.  An all too
brief review of some of the approaches which have had an influence on what
will follow is given below.

\subsection{Static Geometric Models}

\subsubsection{Growth and Form - Mapping One Species to Another}
In 1917, biologist D'Arcy Thompson published a seminal work titled 'On Growth and Form' \cite{DThomp}. In it
he argued that speciation can be understood in terms of geometry, mechanics and basic physical principles. With 
striking visual examples, he shows that simple mathematical transformations lead to variations in 
form that we perceive as different species. Though not outwardly contesting Darwin's ideas, which
were already dominant at the time, Thompson does emphasize that efficient geometries are found everywhere in
the natural world, thereby pointing to principles other than just the survival of the fittest in shaping new
generations of plants and animals. In the 1000 plus pages of his work, Thompson discusses and derives the 
mathematics of a myriad of different phenomena: honeycomb packing, logarithmic spirals of shells, stress lines in bones 
(and how these can be used to build better bridges), cell structure, et cetera, et cetera.

Thompson did not, however, postulate a testable model for his observations. True to the spirit of early 20th
century biologists, his work was to catalogue and describe. Although the word ``growth'' appeared in the 
title of his manuscript, very little discussion of the mechanism for growth was
provided.  However, his underlying motive of understanding 
organisms through the prism of geometry was incredibly
original for its time, and is especially close to the concept we attempt to work with. Indeed, it has already
been a source of inspiration for many students of biomechanics and quantitative biology.

\subsubsection{L-Systems}
L-systems, named after Aristid Lindenmayer, 
are algorithms for describing the fractal nature of
many plants \cite{PL}. They arise from formal grammars where a plant is represented by a string of symbols which signify,
for example, a stem and branches. 

A formal grammar is an algorithm that acts on a set of symbols. From an initial string, each iteration of the 
algorithm generates a new string. For example, one can start with the symbol $S$ and the rule $S \rightarrow qS$, 
which leads to the string $qqqqqqqS$ after seven iterations. For a stem $S$ with branches $b$, one could try
starting with $S$ and the rule $S \rightarrow SbS$. The first three iterations are: $SbS$, $SbSbSbS$, 
$SbSbSbSbSbSbSbS$. 

As can be seen even with a few simple examples, this type of recursive algorithm readily gives self-similar patterns.
The connection to plants is that they often display self-similar, fractal patterns. Indeed, L-systems produce
intricate models of tree branches, simple shrub plants and flowers. With the addition of a few 
parameters to control things like branching angle, rate of growth and the like, a multitude of realistic 
plant models can be produced.

The focus of L-systems, however, is on the large scale architecture of a plant rather than the individual parts 
of a plant. Hence, our chosen approach is complimentary to L-systems and is aimed at the next smallest level of 
plant form. 

\subsection{Fitting to Given Time Dependent Functions}

In many cases, plant growth undergoes three separate phases.  Initially the growth is slow and as the plant 
matures it goes through rapid growth stage, after which it slows to reach a maximum size.  In some cases the
plant will shrink in its old age.  The general shape of a growth curve can be described by a sigmoidal function of time.
Different sigmoidal functions (e.g.~Gompertz, logistic, error, arctan, hyperbolic tangent, etc.) have been used 
to model the growth patterns observed in plant structures. Associated with these functions are various parameters
that hopefully one can relate to actual measurement processes.  However, the functions do not arise out of
a dynamic model, where different variables are evolved from initial data (along with a set of appropriate
spatial boundary conditions).  The functional models provide a means for classifying growth but unfortunately
are unable to provide information about what interactions among the dynamical variables control growth. 

\subsubsection{Observations of Tobacco Leaf Growth}
Since the early 1930s, particular interest has been devoted to documenting the growth and morphology of 
tobacco ({\it Nicotiana tabacum}) leaves \cite {Avery}. The earliest experiments were done by J.S. Huxley who imprinted regularly spaced ink dots on
a maturing tobacco leaf and monitored how these dots moved relative to each other as the leaf grew. This lead 
directly to the ideas of relative growth rates rather than absolute growth rates, and hence the first steps
toward a differential and geometric approach of analysing plants. 

In later decades, a common genetic cause for growth was found for plants, and studied most intensively in tobacco
and arabidopsis plants. Tests on these plants have verified that the growth hormone auxin is responsible
for the rate of plant growth. If genes governing the production of auxin are turned off, the resulting 
plant is stunted to a fraction of its normal height. Conversely, if auxin levels are increased by, for instance,
physically applying more auxin to localized areas on a leaf, that portion of the leaf will grow much faster.

In more recent tobacco leaf research, biologists have been able to describe in great detail the steps of tobacco 
leaf growth by examining it cell by cell. Others have even attempted to characterize the time evolution of certain 
plant parameters (notably the relative growth rate) by fitting regression functions to plant growth data \cite{KI}. Hence, 
the growth of a tobacco leaf is known from a macroscopic scale down to the cellular level, 
but only in a descriptive capacity. This details, however, are of great value in our research because 
they provide trustworthy data to emulate. 

%

\subsection{Edge Buckling due to Mechanical Stresses}
Another recent development on the biomechanical front has been to look at how stresses in a 2-dimensional sheet 
yield buckling patterns along its edges \cite{Derv}. The stresses can be produced either by tearing a sheet while it is under
tension, or by increasing the amount of material on the sheet along its edge. Both approaches yield ruffled patterns
similar to a lettuce leaf edge, and have a fractal structure. Despite visually accurate patterns and control 
parameters that produce a variety of shapes, the basic mechanisms used here explain only the final shape of a surface
and not its development from one surface to another. 

\subsection{Continuum Mechanics Approach}
One of the first attempts at using continuum dynamics to describe plant growth were published by W. Silk and R. Erickson \cite{SE} in 
the late 1970s. Their arguments rested on employing a continuity equation when describing various parameters in 
plants like hormone distributions. Indeed, the experimental data they use shows continuous distributions, especially
when materials are averaged over the whole cell, indicating
that the cells transfer materials between each other rather than concentrating materials more in certain cells 
than others. This is not so surprising - cells have permeable walls and the materials passed between cells are
in a water solution.

The other emphasis in Silk and Erickson's work is on choosing between Eulerian and Lagrangian representations of
the coordinates on which the equations exist. In an Eulerian system, the coordinates are external to the plant 
(i.e. the plant has been put inside a box with, say, a Cartesian system on its axes). In a Lagrangian system, the
coordinate origin is bound to a particular feature of the plant, for example the apex of the stem. As Silk and
Erickson show, the latter representation often reveals stationary material distributions that an Eulerian
system would not show as readily. In terms of what different observers perceive, these results tell us that
an external observer would see materials pushed towards the growing regions of the plant, while the plant 
experiences a constant supply of the materials it needs to grow at a certain rate.

Silk and Erickson did not take their research into the realm of testing predictions, but 
were able to describe certain elements of plant growth in the language of fluid mechanics. Their approach
is the closest to the one proposed in this work; they showed preliminary results that indicate the validity
of fluid dynamics in modeling plant growth. What Silk and Erickson did not consider is how the curved geometry
of the plan leaf can affect the material distributions therein. This aspect will be a main focus of our work.

\section{Constructing A Dynamic Geometric Model}

\subsection{Leaves as Differentiable Manifolds}
The mathematical analysis of a surface requires defining a continuous space of appropriate dimension to the problem
at hand. Once a space, or manifold, has been defined, the tools of differential geometry can be employed to explore 
the properties of that space. 

These tools of differential geometry all stem from considering calculus on a curved space. Of particular interest
for our purposes is correctly defining dynamics - quantities that change over space and time - in curved space.
Essentially, one would like to know how the dynamics of a system are changed because of it's curved geometry, and
how the geometry reacts to the dynamics when the system is set in motion.

A good illustration of this principal is found in general relativity, where the dynamics of particles are affected by the 
curvature of spacetime, but the sources of curvature are massive particles. Hence, there is constant feedback 
between the moving, massive particles and the geometry of the spacetime they move through. 

In the case of plants, the geometry of a leaf or tendril clearly affects the motion and deposition of material through
it. Correspondingly, though, where material (such as water and nutrients) travel to dictates how the plant will 
grow and hence its
geometry.

We cannot, however, omit the fact that plants are inherently discontinuous at the cellular level, whereas differentiable
manifolds are by definition smooth. In this point rests our aforementioned first assumption: that at a macroscopic level, the 
geometry of plants can be described by a smooth geometry. Indeed, it is plain to see that under normal circumstances,
plants do look smooth to our eyes. Moreover, cells are themselves flexible and all but the hardest ones can bend, 
elongate or compress to accommodate the general shape of the leaf or tendril. Finally, in analogy to theoretical
physics, we currently know that at the smallest scales our universe is discontinuous and governed by quantum
mechanics, but we also know that large-scale dynamics are accurately represented by general relativity. Since the
question pursued in this work pertains to the macroscopic geometry of plants, it is not unfounded to apply the
corresponding mathematics of differential manifolds to attempt a solution.

\subsection{Mathematical Preliminaries - Riemannian Geometry}
\label{CF}

In this section the fundamentals of Riemannian Geometry are introduced in preparation for
what follows.  The basic geometrical object in Riemannian geometry is the metric tensor $\mathbf{g}$
with components $g_{ik}(\mathbf{x})$. Given a set of coordinates $\{x^i\}$ ($i=1,n$) that describe points on a $n$-dimensional manifold (geometry), the distances $ds$ between infinitesimal coordinate displacements $d x^i$ are given by:
\beq
ds^2 = g_{ik} dx^i dx^k.
\eeq
where the Einstein summation notation is used such that any pair of repeated indices implies a
summation using that index over the range from $1$ to $n$. It must be remembered that the coordinates 
themselves do not determine the distances between them unless the metric components are exactly one.  
For example one often uses angular coordinates $\theta$ to distinguish between points on a 
circle of radius
$R$. The distance between $\theta_2$ and $\theta_1$ along the circle however is given by $\Delta s = 
R \Delta \theta$ where $\Delta \theta = \theta_2 - \theta_1$. The factor multiplying the difference in the 
coordinates is represented by a metric coefficient.  

As a basic example, let us take flat 2D space with the Cartesian coordinates $(x,y)$. The distance between two points is 
\beq
ds^2 = dx^2 + dy^2,
\eeq
\noi from which we can read the coefficients of the metric tensor to be $g_{11} = 1 = g_{22}$  and $g_{12} = 0 = g_{21}$. Often, the
metric tensor for an $n$-dimensional space is represented by an $n\times n$ symmetric array. The flat space metric is then
\beq
\eta_{ik} = \left(
\begin{array}{cc}
1 & 0 \\
0 & 1
\end{array}
\right).
\eeq

In curved geometries, a non-tensorial quantity called the affine connection is required to
ensure that derivatives of tensor quantities also form tensors. In general, partial derivatives of 
tensor quantities are not themselves tensors. The connection coefficients are objects that
allow one to form tensor derivatives (also called covariant derivatives) through a cancellation of the 
non-tensorial terms appearing in the partial derivatives.  If the connections are determined from
the metric tensor they are often called Christoffel symbols and are are determined from a combination of
partial derivatives of the metric by:

\beq
\Gamma_{bc}^a = \frac{1}{2} g^{ad} \left( \frac{\partial g_{bd}}{\partial x^c} + \frac{\partial g_{dc}}{\partial x^b}
                                + \frac{\partial g_{bc}}{\partial x^d} \right).
				\eeq

To complicate matters even further, the components of vectors and higher order tensors will depend on
whether they are computed with respect to the coordinate displacements (contravariant components), or 
with respect to directions
orthogonal to the coordinate displacements (i.e.~as gradients and known as covariant components).  
The tensor transformation laws for
the components will depend upon how the components are computed, and the derivatives of the tensor
components will depend upon how the the non-tensorial partial derivative terms are removed.

The covariant derivatives of covariant and contravariant components of a vector $\mathbf{A}$ are
given by:

\beq
\nabla_k A_i = \frac{\partial A_i}{\partial x^k} - \Gamma_{ik}^j A_j
\eeq

\beq
\nabla_k A^i = \frac{\partial A^i}{\partial x^k} + \Gamma_{jk}^i A^j.
\eeq
respectively.

On a curved manifold covariant derivatives do not commute (although mixed partial derivatives do). A measure
of the strength of the non-commutativity for a covariant vector is given by the Ricci identity.
$$ \nabla_i \nabla_j A_k - \nabla_j \nabla_i A_k = R^\ell_{kij} v_\ell $$
where $R^\ell_{kij}$ is the Riemann curvature tensor
given by:
$$ R^i_{jk\ell} = \frac{\partial \Gamma_{j\ell}^i}{\partial x^k}  - \frac{\partial \Gamma_{jk}^i }{\partial x^\ell} + \Gamma_{j\ell}^m \Gamma_{mk}^i - \Gamma_{jk}^m \Gamma_{m\ell}^i. $$
Another measure of curvature is provided by the Ricci curvature tensor $ R_{ik} = R^{\ell}_{i \ell k}$
or:
\beq
R_{ik} = \frac{\partial \Gamma_{ik}^m}{\partial x^m}  - \frac{\partial \Gamma_{im}^m }{\partial x^k} + \Gamma_{ik}^n \Gamma_{nm}^m - \Gamma_{im}^n \Gamma_{nk}^m.
\eeq
which can be thought of as a metric weighted average of the Riemannian curvature.  The advantage of
the Ricci tensor is that it has fewer components, is easier to compute and has the same transformation properties as the metric 
tensor.  It does have the disadvantage of sometimes being identically equal to zero even though the Riemann tensor 
components from which it is constructed may be non-zero.

The curvature tensors constructed above are called intrinsic curvature measures since they are computed
only from information contained in the metric components within the manifold's geometry.  There is no
reference to higher dimensional spaces within which the $n$-dimensional manifold might be
embedded.  Therefore whether a geometry is curved or not is something that can be computed from
the properties of the geometry itself. Thus Riemann was able to complete Gauss' programme which was to
provide a method of determining the curvature of a surface (e.g.~that of the Earth) without having
to go off of the surface into a three dimensional embedding space.

Due to a remarkable series of theorems \cite{JN} it has been shown that any Riemannian space can be embedded
into a higher dimensional flat (non-curved) space.  In many cases the dimension of that flat space is $N = n+1$ although there are important exceptions where the dimension of the embedding space may be even higher. 

\subsection{Dynamical Riemannian Geometry}
\label{mathFormalism}

The next step in constructing a dynamical or time evolving geometry is determine the equations 
that govern the time dependence of that geometry and the objects that couple to the geometry.  Once again, Einstein's
theory of general relativity is provides one possibility where the 4-dimensional (pseudo-Riemannian) 
manifold is described by a single time coordinate in addition to three spatial coordinates. Due to
the symmetry between space and time, the general properties of the Einstein equations are that 
they are essentially hyperbolic. This means that gravity can change due to the propagation of waves 
that carry information at a finite velocity from one point in the manifold to another.

On the other hand biological (and chemical) processes 
occur through diffusive flows which are governed by parabolic equations.  Fortunately so-called
curvature flow equations have exactly this this property.  The best known case is the Ricci flow
where the evolution of the metric is determined directly from the Ricci tensor by 
\begin{equation}
\frac{\partial g_{ik}}{\partial t} = \kappa R_{ik}.
\end{equation}
\noi where $\kappa$ is a real constant. Pure Ricci flow (with 
$\kappa - 2$) describes spaces that contract in the direction of positive curvature and expend in the
direction of negative curvature.  

Note that this PDE has both a diffusive and reactive term.
In fact this is no coincidence. In the early 1980's, the mathematician R. Hamilton \cite{Ham} wrote down the equation for
Ricci flow so that manifolds could be evolved through a heat-like process to find out if an arbitrary manifold
is isomorphic to some simpler geometry by smoothing out the irregularities of the arbitrary manifold.
Just as a hot spot in a metal rod will diffuse until the rod is equally heated throughout its length, 
a curved manifold can diffuse its curvature, causing a time evolution of the space that effectively flattens it out.
The motivation for introducing Ricci flows was initially to provide an application to the Thurston geometrization conjecture \cite{Thurst} 
which subsequently led to the resolution of the Poincar\'{e} conjecture \cite {Perel}.

While Ricci flow is a purely intrinsic flow, extrinsic (or mean) curvature flow provides an alternative 
set of differential equations for the metric in terms of the coordinates of the higher dimensional space within which 
the original space is embedded. Assuming that the coordinates of the higher dimensional space are given by $X^\mu(\mathbf{x})$  
(where Greek indices $\mu = 1, \cdots, N$ are used to label coordinates in
the higher dimensional embedding space), the components of $g_{ik}$ can be written as:
$$ g_{ik} = G_{\mu \nu}X^\mu_{,i}X^\nu_{,j} $$
where $X^\mu_{,j} = \partial X^\mu/\partial x^j$ and $G_{\mu \nu}$ is the metric tensor of the embedding space which
is often taken to be the flat space metric, $\eta_{\mu \nu} = {\rm diag}(\pm 1, \cdots , \pm 1)$. 

Following the results of Tapia \cite{Tapia} one can translate the mean curvature flow equations:
$$ \frac{\partial \mathbf{X}}{\partial t} = \nabla^2\mathbf{X}$$
(where the indexed coordinates $X^\mu$ have been replaced by the vector $\mathbf{X}$, and the Laplacian operator, $\nabla^2 = g^{ij}\nabla_i \nabla_j$)
into a flow equation for the metric $\mathbf{g}$:
$$ \frac{\partial g_{ik}}{\partial t}  = \kappa \nabla^2 \mathbf{X} \cdot \nabla_{ij}\mathbf{X} .$$
In the above equation the inner product represented by the $\cdot$ operator is taken with respect to $\eta_{\mu \nu}$.

Curvature flows have primarily remained in the realm of mathematicians, and are just beginning to be applied in 
physical contexts \cite{Woolg}. The 
intuition behind curvature flows for botanical growth seems appealing, though, when one considers that young plants, often 
with tightly 
curved leaves, must somehow develop into mature, flatter leaves; in other words, a plant's geometry and curvature 
are time dependent. A curvature flow process would at once be an elegant and physically motivated application of such a theory.

For most cases leaves and petals when considered on the large scale have negligible thickness and can
therefore be considered as 2-dimensional surfaces.  A 2-dimensional manifold is also \emph{conformally flat}. This means that the metric can be
written in a form where a single function of the spatial coordinates multiplies the 2-D flat metric 
described above. This is consistent with the fact that the Riemann curvature tensor in 2-dimensions has only one
independent component. Thus introducing a set of two spatial coordinates $(u,v)$ the metric on any 2-D structure
can be written in terms of a single  scale factor $f(u,v;t)$.
\beq
g_{ik} = f(u,v;t) 
\left(
\begin{array}{cc}
1 & 0 \\
0 & 1
\end{array}
\right).
\eeq
\noi The fact that $f$ is time dependent indicates that
the distances between points on the manifold can change over time. Clearly, working with a single function 
rather than three independent parameters is always advantageous, and will prove to be so in subsequent calculations.

The Ricci tensor describing the curvature of this manifold at any moment in time also has one independent component
given by:
\beq
R_{11} = R_{22} = \frac{\kappa}{2f} \left[ \frac{\partial^2 f}{\partial u^2} + \frac{\partial^2 f}{\partial v^2} - \frac{1}{f} \left[ \left(\frac{\partial f}{\partial u}\right)^2 + \left(\frac{\partial f}{\partial v}\right)^2 \right] \right]
\eeq

Just as a 2D space will be used when considering leaves, a 1D space can be used to represent roots and tendrils, and
spaces that have a symmetry in one of their two coordinates. 
The metric tensor for a 1D manifold is even simpler than in the 2D case as there is only one coordinate axis and
only one parameter that describes how the manifold is shaped:
\beq
g_{11} = f(u;t).
\eeq

\subsection{Coupling Material Flow, Growth and Geometry}
\label{theModel}

The basic premise on which Silk and Erickson based their study of plant growth was that material flow in the leaf should
be governed by the laws of material transport. This makes sense since the leaf relies on the flow of 
nutrients and growth hormones to each cell in order to grow. The cells in turn create more cells through
cell division and this accounts for the increase in spatial dimensions. Furthermore, the growth is symplastic. That is the 
cell growth and formation is smooth and coordinated in such a away that the
cells once formed do not migrate. This makes the continuum description easier
than in the case of animal tissue growth where the cells can slide over one
another. The equations governing tissue growth should
account for the flow of material throughout the medium as well as the change in
distances between fixed coordinate locations on the body.  It is therefore
natural to impose a metric whose coefficients depend on both space and time.  

Since plant growth is an example of an open,  non-equilibrium dynamical system,
we apply the philosophy presented in Kardar's text \cite{Kard} that the {\it equation
of motion} is the fundamental object of interest in describing systems that
lack a Hamiltonian formulation.  The equations of motion should include all
terms consistent with the symmetry of the problem since ``in a generic situation
an 
allowed terms is present and only vanishes by accident''. Furthermore for a spatio-temporal field 
$h({\mathbf{x}},t)$ one can expect that in the presence of dissipative dynamics and over long
enough time scales inertial terms (i.e. those proportional to $\partial_t^2h$) are irrelevant.

In general the evolution of $h$ is governed by an equation of the form:
$$ \partial_t h({\mathbf{x}},t) = f[h,{\mathbf{x}},t] + \phi[{\mathbf{x}},t] $$
where the first term on the RHS governs the deterministic dynamics and the
second introduces stochastic terms. If the interactions are short ranged
the driving terms in the function $f$ at $({\mathbf{x}},t)$ depends on $h$ and a few
low order spatial derivatives:
$$  f[h,{\mathbf{x}},t] =  f[[h({\mathbf{x}},t),  \nabla h({\mathbf{x}},t),
\cdots {\mathbf{x}}, t] $$

Therefore we can expect a feedback system where the flow of material during
growth creates stresses that change the leaf's geometry which in turn affects the
rate of flow of material.                          
 
Once again, this is not that far removed from the ideas of general relativity. In the context of spacetime, Einstein 
postulated that the movement of massive objects such as planets, stars and galaxies is governed by the shape of the spacetime 
through which they move, but that the geometry of spacetime is shaped by massive objects residing in it. 

When the plant leaf is viewed as a curved 2-dimensional space, then it is natural to think of the material flowing
on it as reacting to the curvature it must navigate, just as we react to the curved spacetime around our planet by 
experiencing gravity. This aspect is, however, missing from Silk and Erickson's work, in which they applied fluid
flow like models based in a flat background space.

In order to incorporate curvature effects into a fluid flow model, we must translate our assumptions into valid 
differential equations involving tensors quantities, such as scalars, vectors and higher rank tensors. Tensor
equations, when properly formulated, are invariant with respect to arbitrary
coordinate choices and therefore provide fundamental relationships between
dynamical variables.  By extending the theory of Silk and Erickson we introduce
three dynamical variables, (1) the metric tensor which describes the geometry
of the curved manifold that makes up the biological organ, (2) the material flow
velocity of the material that causes the growth of the organ itself and (3) the
density of the material that makes up the organ.  Each variable will depend on
both spatial and temporal coordinates and equations of motion will be developed
tin a manner following the guidelines discussed in the preceding paragraphs.    

The equation for the the evolution of
the metric tensor, describes changes in the leaf's geometry in response to the stresses induced by the
flow of material in the leaf.  In addition the curvature of the leaf may also
induce changes in the metric tensor. 
A vector equation for the velocity field of the material 
flowing on the leaf should include diffusive, advective, damping, pressure and external force terms, as well as 
a reaction of the flow to the leaf's geometrical structure.
Finally, a scalar equation for the density of the material that makes up the leaf can be written should be
in the form of a continuity equation having both sources and sinks. If tensorial equations are to have an intrinsic meaning,
they must be independent of the the coordinates chosen to describe of measure the components of the objects in question.
Therefore we will develop a set of dynamical equations where the variables in the equations all obey the
same transformation rules.  Such a restriction requires the use of self-consistent expressions not only in the actual
dynamical variables but also in how their derivatives with respect to spatial coordinates appear in the equations.

First, let us construct the equation for the evolution of the metric. We would like the geometry of the leaf to respond 
to the flow of
material. In other words, a change in the flow of material should affect the time evolution of the geometry:

\beq
\frac{\partial g_{ik}}{\partial t} \propto  \nabla_i v_k + \nabla_k v_i.
\eeq

\noi In biology, $\partial v^k/ \partial x^i$ (or in this case its covariant equivalent $\nabla_i v_k$) is called the growth tensor \cite{Naj}, and can be measured experimentally. It is a measure
of how the velocity field on the leaf varies spatially. 
In continuum mechanics,  $\gamma_{ij} = 1/2(\partial v_i/\partial x^j + \partial v_j/\partial x^i$) is the deformation tensor
or in fluid dynamics the strain rate tensor.  Since
a Riemannian metric tensor $g_{ik}$ is itself symmetric, a symmetrized form of the growth tensor (i.e.~the strain rate or
deformation) is
introduced to account for this fact.  From a dimensional analysis point of view the time derivative of the metric has units 
of inverse time, exactly those units for the growth tensor. 

Likewise, the curvature of the plant should be able to change over time, changing from a highly curved young leaf
to a more flat mature leaf. This is where curvature flow affects the geometry
and this represents a nonlinear feedback of the geometry back onto itself:  

\beq
\frac{\partial g_{ik}}{\partial t} \propto R_{ik},
\eeq

\noi where $R_{ik}$ is the Ricci curvature, as discussed in Section \ref{CF}.  

The final ingredient in the geometric evolution is the possible presence of a nonlinear stress:

\beq
\frac{\partial g_{ik}}{\partial t} \propto v_i v_k.
\eeq

Taken together, these proportionalities form the most generalized set of tensorially consistent quantities
involving all the possible second-rank tensors under consideration. The metric tensor time evolution
is therefore governed by 

\beq
\frac{\partial g_{ik}}{\partial t} = \kappa R_{ik} + \kappa_1 v_i v_k + \kappa_2 [ \nabla_i v_k + \nabla_k v_i],
\eeq

\noi where $\kappa_i$ are real constants reflecting the strength with which each term affects the evolution.

We now turning our attention to the vector equation for the leaf's velocity field. Most of the terms are 
borrowed from standard fluid mechanics and represent, respectively, advection, damping, external forces,
pressure gradients and diffusion.  From this point of view the model is similar
to that introduced by Silk and Erickson:

\beq
\frac{\partial v^i}{\partial t} \propto  -  v^k \nabla_k v^i + \frac{1}{\rho} [ v^i +  F^i
						+  g^{ik} \nabla_k P ] + g^{lm} \nabla_l \nabla_m v^i.
\eeq

The additional term needed to have the flow respond to curvature is introduced by a ``force'' representing
motion along the geodesics of the geometry. The introduction of geodesic flow appears not only in
general relativity but in condensed matter physics where they have been found experimentally to
be related to the strain tensor patterns observed in sheets of disordered material \cite{Jeulin}. The non-linear
(quadratic)
coupling of the flow velocity with itself, together with the viscous,
dissipative terms above provides a reaction-diffusion type equation for the
velocity field: 

\beq
\frac{\partial v^i}{\partial t} \propto -\Gamma_{jk}^i v^j v^k.
\eeq

Together, the time evolution of the velocity field reads

\beq
\frac{\partial v^i}{\partial t} = -c\Gamma_{jk}^i v^j v^k -  c_1 v^k \nabla_k v^i + \frac{1}{\rho} [ c_2 v^i + c_3 F^i
						+ c_4 g^{ik} \nabla_k P ] + c_5 g^{lm} \nabla_l \nabla_m v^i 
\eeq

\noi where $c_i$ are real constants.

Lastly, a scalar equation can be written to describe changes in the leaf's density over time as a result of 
sources (or sinks) represented by the function
$S({\mathbf{x}}, t)$  and flux of material flowing into or out of the a closed
region of the leaf:

\beq
\frac{\partial \rho}{\partial t} = S({\mathbf{x}}, t) - \nabla_i(\rho v^i) 
\eeq

It is important to note that in all the equations presented above, the use of $\nabla_i$ indicates a covariant
derivative as presented in section \ref{mathFormalism}. This is yet another way that geometry affects the evolution of 
the dynamical variables describing the system.

\section{One-Dimensional Plant Growth}

In considering a geometric model of plant growth, it would be advantageous to study a simple system first,
and then proceed to build a more general model. This can be done by looking at growth in one-dimensional. 
This is not without immediate application. For example the growth of 
stems, blades of grass, and tendrils might well be approximated by one-dimensional systems.
Similarly plants with circular symmetry, such as the lotus leaf, can have
geometries described by a single radial coordinate.

Reducing the system to 1 spatial dimension and a 1-dimensional velocity field eliminates the possibility
of finding spatio-temporal chaos, as the parameter space of a chaotic system must be at least 3-dimensional.
Hence, a 1-dimensional plant promises to be a well-behaved system, which is advantageous when performing
numerical simulations on a system so poorly understood. Also, building intuition about the simpler
scenario becomes all the more important when moving into the 2-dimensional model.

\subsection{Assumptions and Approximations} \label{approx}

One-dimensional geometries have a Riemann-tensor that vanishes, i.e.~1D
systems have no intrinsic curvature.  This means that the nonlinear feedback
that would normally appear in coupling the Ricci tensor to the metric now
vanishes identically and the evolution of the metric would be governed directly
by the fluid flow.
 
In order to provide a nonlinear coupling between the scale factor and its
derivatives with respect to the spatial coordinates, we introduce a
``pseudo-curvature'' term that contains terms linear in the Laplacian of the
scale factor and quadratic in the first spatial derivatives.  This will act as a
substitute for the Ricci tensor in the 1D case. In addition to this, it is
assumed that the leaf consists of a constant density material.  This seems to
be a reasonable assumption based upon experience and clearly indicates that the
flux of material is then driven by an external source that sustains the growth,
at least for early times. In addition external forces such as gravity and
imposed pressure gradients will also be ignored in favour of focusing more on
the internal dynamics.

Choosing the spatial coordinate
to be $x$, the 1D metric scale factor $f(x,t)$ the flow velocity $v(x,t)$ and
the material density $\rho(x,t)$ obey the following equations: 

\begin{eqnarray}
\frac{\partial f}{\partial t} & = & \frac{\kappa}{2f} \left[ \frac{\partial^2 f}{\partial x^2} - \frac{1}{f} \left(\frac{\partial f}{\partial x}\right)^2 \right] + \kappa_1 f^2 v^2 + \kappa_2 \left( 2 f \frac{\partial v}{\partial x} + v \frac{\partial f}{\partial x} \right)\\
\frac{\partial v}{\partial t} & = & -\frac{c_0}{2f} \left(\frac{\partial f}{\partial x}\right) v^2 - c_1 v \left[ \frac{\partial v}{\partial x} + \frac{1}{2f} \left(\frac{\partial f}{\partial x} \right) v \right] - c_2 v \nonumber \\
& &  - c_3 \left[ \frac{1}{f} \frac{\partial^2 v}{\partial x^2} + \frac{1}{2f^2} \left( \frac{\partial^2 f}{\partial x^2} v + \frac{\partial f}{\partial x} \frac{\partial v}{\partial x} - \frac{1}{f} \left( \frac{\partial f}{\partial x} \right)^2 v \right) \right].
\end{eqnarray}

Admittedly, these equations, even with the simplifying assumptions are rather complex. However there are situations
where some well known dynamical systems can be recovered from the general
dynamics in the appropriate limiting cases. 

For example the velocity equation (when $c_1=c_2=0$) reduces to a reaction-diffusion equation for the velocity:

\beq
\frac{\partial v}{\partial t} = -\frac{c_0}{2f} \left(\frac{\partial f}{\partial x}\right) v^2 - c_3 \frac{1}{f} \frac{\partial^2 v}{\partial x^2} + \cdots .
\eeq 

\noi When $f=1$, and $c_0 = c_3 = 0$, $c_1 = 1$) one obtains Burger's equation \cite{Burg}
if $-c_3$ is identified as a viscosity parameter:

\beq
\frac{\partial v}{\partial t} = - c_1 v \frac{\partial v}{\partial x} - c_3 \frac{\partial^2 v}{\partial x^2}.
\eeq

\noi This is not altogether surprising, given the relationship between these equations and those describing fluid flow. The
coupling to a dynamical background geometry is what introduces the largest
number of terms in the equations. 

Intriguingly though, the metric tensor evolution has hints of a process not originally inscribed in its
architecture. The so-called pseudo-curvature term that replaces the Ricci tensor in the 1-D case
 reduces to a Kardar-Parisi-Zhang (KPZ) type equation \cite{KPZ}. The KPZ system was designed to study rapid crystal growth
(particularly from the deposition of material) and is an evolution equation for
the surface height $h({\mathbf{x}},t)$ where $x$ is a coordinate measured along
the boundary between the surface and the air (or vacuum).
$$ \frac{\partial h({\mathbf{x}},t)}{\partial t} = \nu \nabla^2 h +
\frac{\lambda}{2} (\nabla h)^2 + \cdots .$$

In our case writing the scale factor as a perturbation of flat space $f \approx 1 + h$ and setting
$\kappa_1 = \kappa_2 \approx 0$  leads to:

\beq
\frac{\partial h}{\partial t} \approx  \frac{\kappa}{2} \left[ \frac{\partial^2 h}{\partial x^2} - \left(\frac{\partial h}{\partial x}\right)^2 \right] + \cdots .
\eeq

\noi What in 2-dimensional space is associated with the Ricci flow, in 1-dimension becomes similar to physical
processes normally associated with growth processes. While;e the structure of
the equations is similar, they describe different growth dynamics.  The KPZ
system describes growth perpendicular to the coordinate $x$ whereas our
equation describes a continuous growth in the direction of the $x$-coordinate.

An obvious  steady state solution can be found by letting $\partial f / \partial t = 0$ and $\partial v / \partial t = 0$. 
In that case, 

\begin{eqnarray}
f & = & C_1 + C_2 e^{-bx} \\
v & = &  0
\end{eqnarray}

\noi where $C_1$, $C_2$ and  $b$ are real constants. Physically, this would correspond to a leaf that has 
stopped growing since the material velocity goes to zero, but the leaf has a non-zero length.


The evolution equations at hand can also be studied numerically. Having both spatial and temporal aspects, a
finite differencing scheme must be chosen carefully so as to achieve accurate results as well as long-term
stability. We choose to employ, a semi-implicit Dufort-Frankel scheme for the Laplacian operator since it
 provides a means for using larger time steps than an explicit scheme while
maintaining a numerically stable evolution. 
It is most often used in heat-type equations; for the heat equation in particular,
the Dufort-Frankel scheme is as follows:

\beq
\frac{ u_j^{n+1} -  u_j^{n-1}}{2 \Delta t} = \alpha \frac{u_{j+1}^n - u_j^{n+1} - u_j^{n-1} + u_{j-1}^n}{(\Delta x)^2},
\eeq

\noi where $t = n \Delta t$ and $x = j \Delta x$ and $\alpha$ is a constant of proportionality.

The Dufort-Frankel method leads to an explicit scheme in calculating the time-evolving values of a function and is robust
enough to allow time and spatial differencing values to be comparable when $\alpha \sim 10^{-2}$.

\subsection{Results and Discussion}
\begin{figure}[hbt]
\begin{center}
\scalebox{1.00}{\includegraphics{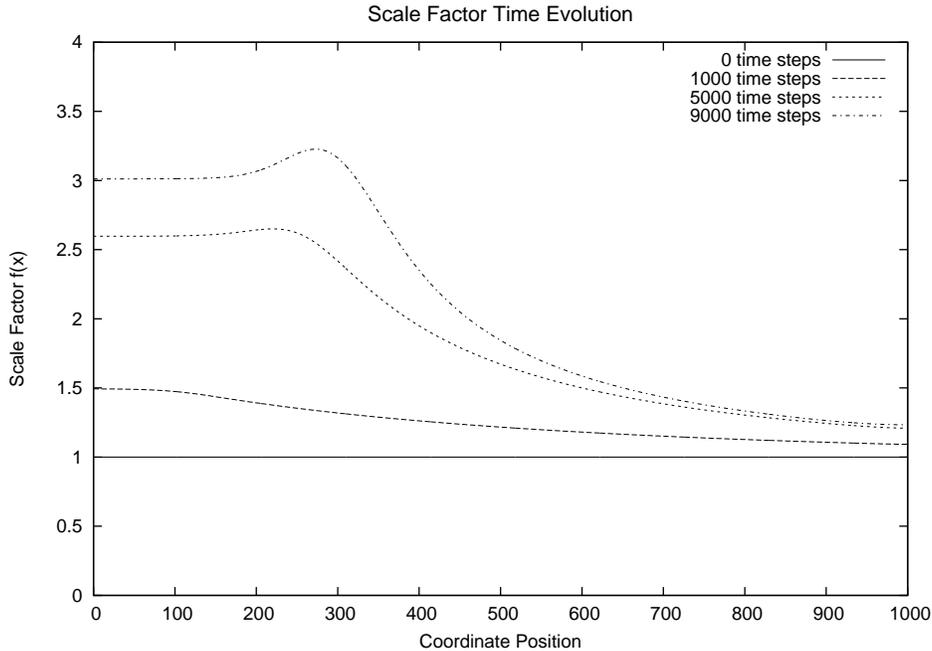}}
\end{center}
\caption{ Plots of the evolution of the scale factor $f(x,t)$.  The plots
are ``snapshots'' (at specified values of $t$) of the spatial distribution 
of the scale factor as
a function of the coordinate position $x$.}
\label{scaleplot}
\end{figure}
\begin{figure}[h]
\begin{center}
\scalebox{1.00}{\includegraphics{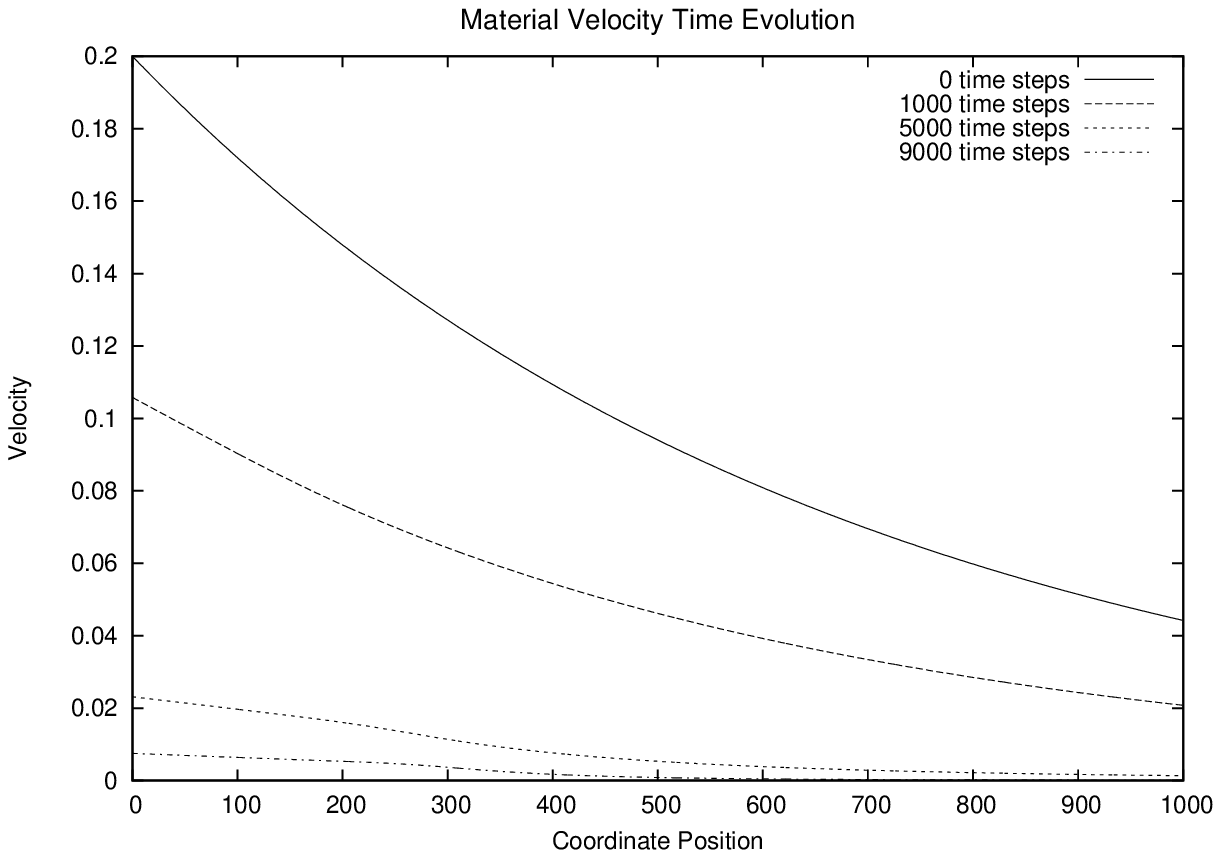}}
\end{center}
\caption{ Plots of the evolution of the velocity $v(x,t)$.}  
\label{velocplot}
\end{figure}

The generic behaviour of the scale factor for all cases is that the scale
Numerical solutions to the equations were constructed for a wide range of parameter values. This allowed one to
emphasize different effects to determine which terms are dominant and which one
can regulate the growth patterns.  In addition a variety of initial conditions
for the spatial distribution of the  scale factor and the flow velocity were
studied.

In what follows we present some results for evolutions obtained with simple initial conditions given by 
a scale factor set to unity every where and a velocity profile that is an 
exponentally decreasing function of spatial position.  The following values for the constants $\kappa_n$ and $c_n$
$$ \kappa = -0.05, \;\;\;\;\;\; \kappa_1 = 0.02,\;\;\;\;\;\; \kappa_2 = -0.08,$$
$$ c= 0.1, \;\;\;\;\;\; c_1 = 0.01, \;\;\;\;\;\; c_2 =c_3= c_4 = 0 \;\;\;\;\;\; c_5 = 0.0001 $$
The evolution for both the scale factor and the material velocity are
shown in Figures 1 and 2 respectively.  

The generic behaviour of the scale factor for all cases is that the scale
factor grows in time, as it should. This means that the distances between the
fixed points on the plant increase with time.  It is interesting to notice
that the early in the evolution the scale factor is a monotonically decreasing function of
position. At later times the scale factor grows in the intermediate region,
indicating that the growth occurs most noticeably in this region.  This is
exactly what has been observed in the spatial distribution of growth patterns
observed in 1D systems such as blades of grass ({\it Festuca arundinacea})\cite{Schny} and the primary root of
corn ({\it Zea mays})\cite{Sharp}.  

The velocity distribution of is shown in Figure 2 where initially the velocity
is a decreasing function of position and as the growth occurs (i.e.~the system
undergoes an increase in length) the speed at which the material is transported 
decreases to zero.  While it would appear that these figures indicate that there is no back-reaction
effect on the plant due to the change on scale factor, simulations that de-couple the
curvature from the evolution (i.e.~an evolution for
$f(x,t) = 1$, for all $x$ and $t$) leads to
unbounded growth.  The effect of stretching and curvature (or in this case the nonlinear KPZ
term) leads to a regulation of the overall increase in length. 

Overall, it is easy to find sets of parameters for which growth is initially rapid, then tapers off as the
plant ages, and finally stops growing. This correlates well with patterns seen in tobacco leaf and other plant
growth. In order to compute the overall growth of the system one needs to
define the length of the system in terms of the scale factor:
$$ L(t) = \int_0^{x_{\rm{max}}} f^{\frac{1}{2}}(x,t) dx. $$
Figure 3 shows how the total size of the system changes as a function of time for two different choices of
spatial boundary conditions.  In both cases presented here
there is a
rapid growth phase that eventually dies out when some maximum length is
reached.  At late times the transport velocity goes to zero everywhere along the structure.  
Not only does the the geometry regulate the growth but changes in the boundary conditions
placed upon far tip of the structure changes the long term growth pattern.  If
the boundary condition is a Neumann condition where the derivative of both $f$
and $v$ vanish, then the growth leads to a constant late time length. On the
other hand a Neumann condition where the derivative is a given small non-zero negative
value, leads to a small and slow shrinking of the overall length as a function of time.
\begin{figure}[h]
\begin{center}
\scalebox{1.00}{\includegraphics{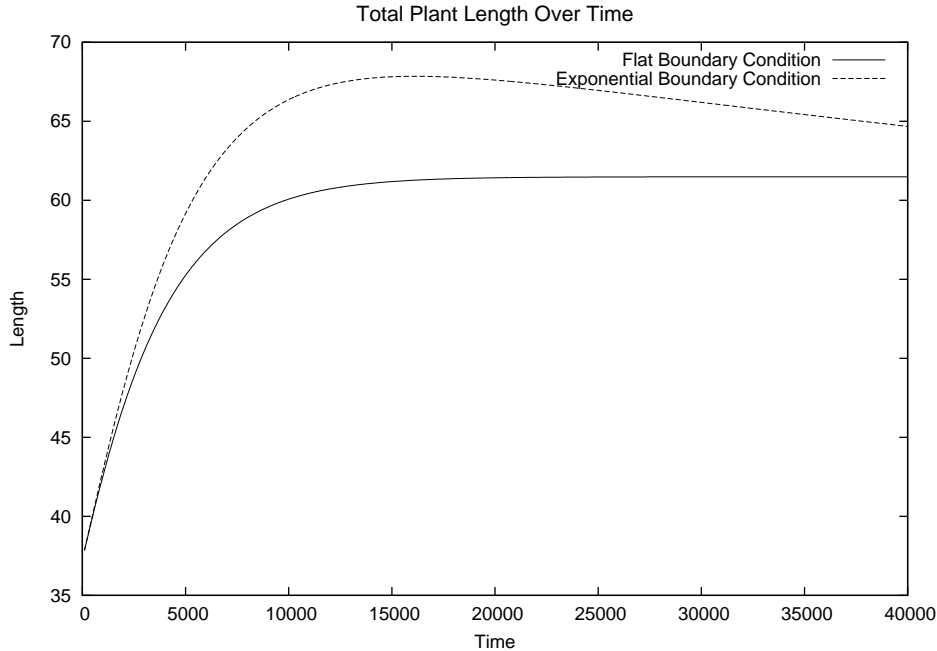}}
\end{center}
\caption{ Plots of: The behaviour of the overall length $L$ as a function of
time.  In both cases the initial conditions are the same ($f(x,0) = 1$, $v(x,0)
= v_0 e^{-alpha x}$) but the spatial boundary conditions are different and this
leads to two different patterns of growth at both intermediate and late times.}
\label{lengthplot}
\end{figure}

Figure 4 which plots fixed coordinate positions as a function of time
demonstrates that the largest changes early in the evolution occur in the region where
the velocity was the largest.  The growth in the outer region is smaller and
the largest contributions to the overall length come from the inner region.
This corresponds exactly to patterns one observes in actual plant growth.  This
figure can be compare directly with Figure 3 appearing in \cite{Sharp} where fixed
positions on a a primary root of a corn plant are indicated by stripes painted
on them.  Similarly the coordinate locations on blades of grass are measured
from pin-pricks made on the young blades \cite{Schny}. Both observations show that the
distribution of growth as a function of position along the organ takes the form
of an inverted ``U''.  The region where the maximum segmented growth appears is
often called the ``elongation zone'' and this is exactly the behaviour one sees develop in the
scale factor evolution shown in Figure 1.    

\begin{figure}[hbt]
\begin{center}
\scalebox{1.00}{\includegraphics{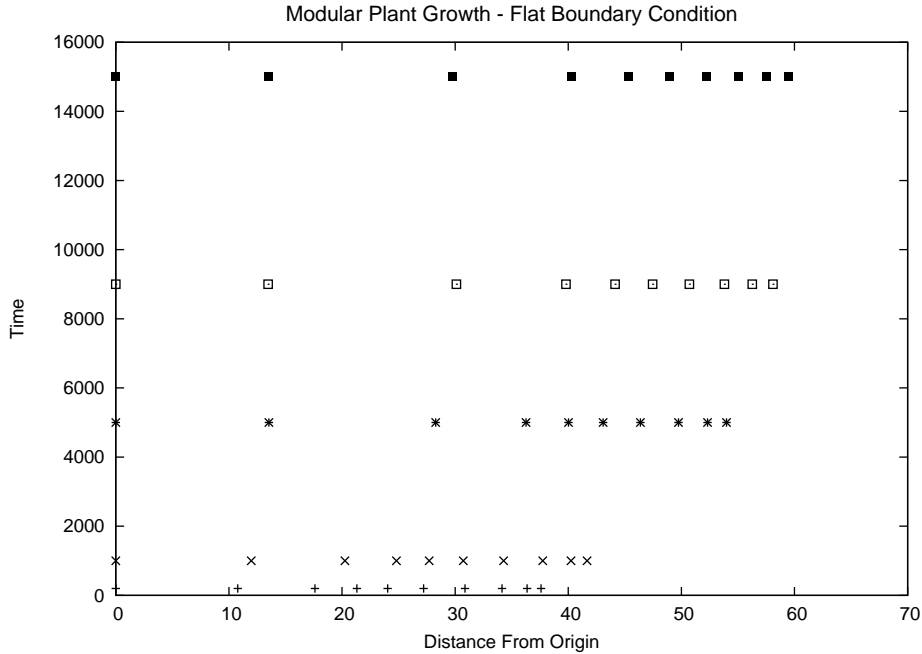}}
\end{center}
\caption{ Plots of the spatial distribution of growth occurring at
specific intervals over the entire length of the plant.  The marks represent
the constant coordinate positions on the plant. This corresponds to experimental methods that 
make permanent marks on a growing plant that are used to monitor growth over a period
of time.}
\label{growthplot}
\end{figure}

What is interesting is to plot the growth tensor profile at different times in
the evolution (See Figure 5).  It can be seen that a local peak in the growth tensor forms
in the inner region of the plant and that the peak ``propagates'' outward 
as the structure grows.  Eventually this peak in the growth tensor will
dissipate as the velocity begins to vanish everywhere at late times.  This phenomena
is consistent with experiments on linear growth where the elongation zone seems
to move with wave-like behaviour outward during the growth process.\cite{Sharp}. 
The formation of the 
such a disturbance from monotonic initial data and its eventual ``propagation''
through the system is a phenomenon that needs more careful study.  

\begin{figure}[hbt]
\begin{center}
\scalebox{1.00}{\includegraphics{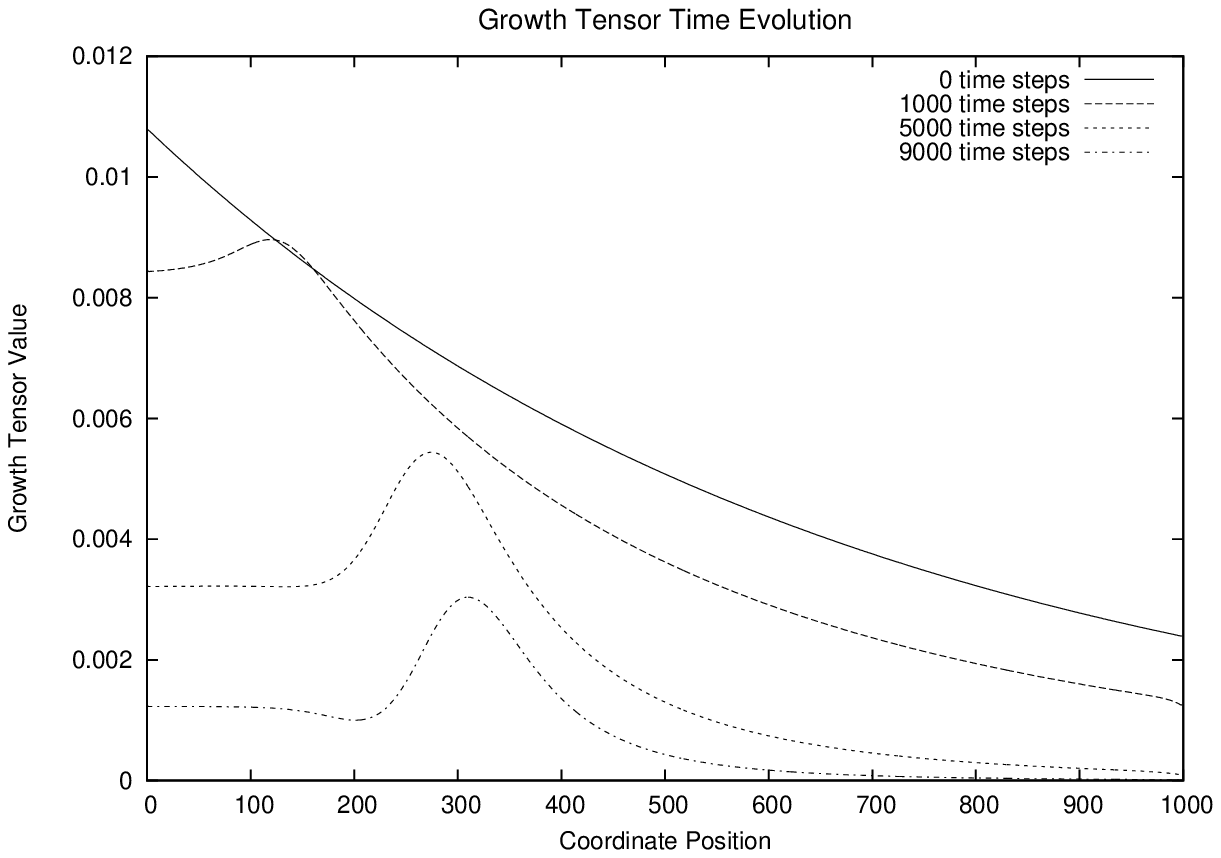}}
\end{center}
\caption{ The evolution of the symmetrized growth tensor $\gamma_{ij}(x,t)$.  
(at specified values of $t$). The peak in the growth tensor is seen to move 
into the elongation zone in a manner reminiscent of a propagating wave.}
\label{gtensorplot}
\end{figure}

Clearly the model presented in this manuscript is able to reproduce many of the 
features one expects from plant growth.  The advantage of using a time dependent metric
is that its use can be directly related to the types of experimental measures one makes
in studying plant growth.  Accepting the fact that the specification of coordinates 
is fixed on the plant once the points on the plant are marked, the metric tensor then
becomes the most natural way to turn the coordinate displacements into distance measurements
especially on a curved surface.

The relationship between the metric scale factor and the behaviour of other quantities 
introduces a dynamical system where the material transport and the change in scale 
are intimately related through nonlinear reaction-diffusion type processes.  This too 
is what one would expect from a time dependent system where diffusive processes
dominate the dynamical evolution.

The results shown above, while encouraging, will require a great deal more effort if
one is to understand the evolution of geometry coupled to growth.  A more quantitative 
analysis of the parameters, along with the initial and boundary conditions will be performed
in order to link the simulations directly to the measurements.  This should help provide
a means for understanding what exactly the parameters describe.  

As with many nonlinear PDEs, the boundary conditions imposed on the system are essential to its behaviour. 
With a simple reflective boundary, the system evolves towards a flat steady state solution, while with an
exponential function fit at the boundary, the metric tensor achieves an exponential steady state. This
corresponds to the analytic solution explored in Section \ref{approx}, but it is unknown whether one solution
is more biologically relevant than the other.

Collectively, the results indicate that geometry plays an active role in material transport across the system,
and hence the plant's growth. By influencing the flow rate of nutrients and hormones, the geometry of the 
leaf can produce unbounded growth, impede growth altogether, or regulate it in a nontrivial fashion.

\section{Future Work}
Numerical studies of the 1-dimensional plant model have provided promising results regarding the role of geometry
in growth, but there are aspects of the model that remain poorly understood. These are largely tied to a lack of
knowledge about the biophysics of plant growth. 

Both the initial conditions and the boundary conditions are, at this point, assigned with no basis in biology, 
in part because biologists do not give these types of questions direct consideration, and therefore there is
no standardized data for these quantities. There are, however, indirect clues in certain biomechanical studies 
of plant growth, and further work will include distilling these studies to better understand the physics behind 
how a plant starts its growth and how nutrients flow at the base and edges of leaves. 

Another aspect that requires attention is the meaning of the various parameters included in the equations. Ideally
these parameters would reflect gene expressions in the plant itself; this would allow for a model in which 
varying the parameters corresponds to modeling different species of plants. Lacking direct evidence for such 
a correlation though, statistical sampling of the parameters may be an adequate method to decide plausible values
for physically realistic plants.

There is also a somewhat philosophical consideration that acts as a hidden assumption in this model. In discussing
the role of curvature and curvature flows, it is important to decide whether one considers the extrinsic curvature
or the intrinsic curvature of the space. For the plant, this translates to deciding whether the plant knows it
is growing in 3 dimensions or 2.

Mathematically, the two approaches have different strengths. In the extrinsic approach, we are guaranteed that
simulations will produce physically realizable shapes, or else fail as soon as a nonphysical shape is encountered.
The trouble is in keeping track of that shape because extrinsic curvature is a scalar-valued quantity and so
cannot be directly incorporated into the time evolution calculations of the metric tensor.

On the other hand, the current approach of intrinsic curvature evolves the dynamics in a lower dimensional space, 
and treats the problem of what the leaf looks like separately. Unfortunately, embedding problems like this one 
are difficult to solve and have the potential to produce ambiguous or unphysical shapes.

Deciding this problem in 1-dimension may prove very helpful before venturing into the more complicated 2-dimensional
model. Indeed, the very dynamics of the two approaches may well be different and in this way provide necessary information
on which is the more physically relevant.

\section{Conclusion}
A dynamical model that couples a non-flat Riemannian geometry to physically
measurable dynamical variables has been developed in order to understand both
the increase in size and the formation of structure in growing plants. The use
of a metric with a scale factor that depends upon both space and time leads
naturally to a system that changes its scale and at he same time has leads to
non-zero curvature.  

In many dynamical systems derived from a least-action principle, one
introduces all the terms that are allowable into the expression for the
Lagrangian and/or Hamiltonian and then obtains the dynamical equations of
motion from standard procedures.  In the case of systems far from
equilibrium, one must introduce the equations of motion directly again
incorporating all allowable terms.  This process seems to have met with success
in this case and or one-dimensional toy model does capture in a general way
many of the properties associated with biological growth.  

The fact that the geometry plays a role in regulating the growth and leads to 
an asymptotic solution at late time is encouraging evidence that the geometry
of a system may indeed affect the growth and vice versa.

What values of the coupling parameters correspond to directly to real plant
observations remain to be explored.  At least for 1-D models there seems to
be very little variation in the overall pattern for growth in changes in 
the parameters.  There is still room for exploration in the 1-D cases that
should lead to even more realistic evolution.

One change is to alter the Ricci flow to mean-curvature flow. The mean
(extrinsic) curvature of a 1-D structure does not vanish as it does for the
Ricci (intrinsic) curvature.  The fact that the extrinsic curvature can written
explicitly was a function of the spatial coordinate (it is a curve embedded in 
a 3-D flat space) makes it easier to specify the initial conditions.  The flow
equation for the scale factor of the metric can still be computed but now it 
is in terms of the equation for the curve.  This has the advantage also that the 
curve can be easily embedded in a higher dimensional flat space explicitly and
makes for easier and more intuitive visualization.  

Secondly the 1-D case can be extended to 2-D spaces with a high degree of
symmetry.  Examples of lotus leaves that are circular will have intrinsic
curvature.  Similarly the evolution of {\it Acetabularia} algae with a circular cap that
undergoes and evolution from a positively curved cap to one that is flat to one
with negative curvature poses the challenge of understanding what it is 
that causes the change in the sign of the curvature. 

Thirdly what role does the continuity equation play in the evolution? If the
density of the structure is not constant how does that effect the overall
growth of the system. This adds another degree of freedom to the system and
that alone should induce some interesting dynamics.    

Finally full 2-D simulations of leaves and petals are likely to show a variety of
different behaviours.  A system with three (or more) dynamical variables will
be more difficult to simulate but pattern formation, spatio-temporal chaos, 
and and other strongly nonlinear phenomena might be expected to occur as they do in other higher dimensional 
systems.   In addition the boundary conditions become more complex.  Since the
simple spatial boundary conditions in the 1D model affect the over all evolution one can expect even more 
complicated situations where the spatial boundaries themselves are changing.

\section{Acknowledgements}
The authors would like to thank a number of people for their interest and
encouragement while this project was being developed. In particular DH would like to thank
Andrew Berdahl and Elliot Martin who as students in the ``Bytes for Life'' class taught by Przemyslaw Prusinkiewicz
were searching for a project that would couple geometry and biological systems. We would also
like to acknowledge the encouragement we received from Maya Paczuski, Joern Davidson, Stuart Kauffman and Leon Glass
during the infancy of this project. 
Finally it was the numerous and fruitful discussions we had with Przemyslaw Prusinkiewicz, Pierre Barbier de Reuille,
Brendan Lane and other members and visitors to the University of Calgary's Biological Modelling and Visualization group 
in the Department 
of Computer Science that that helped us enormously in creating the model described in this manuscript.

This work was supported financially by graduate scholarships from the Canadian National Sciences and Engineering Research Council
and the Alberta Ingenuity Fund awarded to JP.

\section{References}

\end{document}